\documentclass[twocolumn,aps,prb,epsf,showpacs,superscriptaddress]{revtex4}
\usepackage{amsmath}
\usepackage{amsfonts}
\usepackage{epsfig}
\usepackage{epsf}
\usepackage{array}

\begin{document}

\title{Dynamical electronic nematicity from Mott physics}

\author{S. Okamoto}
\altaffiliation{okapon@ornl.gov}
\affiliation{Materials Science and Technology Division, Oak Ridge National Laboratory, Oak Ridge, Tennessee 37831, USA}
\author{D. S\'en\'echal}
\affiliation{D\'epartment de Physique and RQMP, Universit\'e de Sherbrooke, Sherbrooke, Qu\'ebec, Canada J1K 2R1}
\author{M. Civelli}
\altaffiliation{Present address: Laboratoire de Physique des Solides, Univ. Paris-Sud, CNRS, UMR 8502, F-91405 Orsay Cedex, France.}
\affiliation{Theory Group, Institut Laue Langevin, 6 rue Jules Horowitz, 38042 Grenoble Cedex, France}
\author{A.-M. S. Tremblay}
\affiliation{D\'epartment de Physique and RQMP, Universit\'e de Sherbrooke, Sherbrooke, Qu\'ebec, Canada J1K 2R1}
\affiliation{Canadian Institute for Advanced Research, Toronto, Ontario, Canada}

\begin{abstract}
Very large anisotropies in transport quantities have been observed in the presence of very small in-plane structural anisotropy 
in many strongly correlated electron materials. 
By studying the two-dimensional Hubbard model with dynamical-mean-field theory for clusters, 
we show that such large anisotropies can be induced without static stripe order 
if the interaction is large enough to yield a Mott transition. 
Anisotropy decreases at large frequency. 
The maximum effect on conductivity anisotropy occurs in the underdoped regime, as observed in high temperature superconductors.
\end{abstract}

\pacs{74.20.-z, 71.10.-w, 74.25.Jb, 74.72.-h}
\maketitle




Nematics are translationally invariant but spontaneously break rotational symmetry. 
They have been observed long ago in the field of liquid crystals. In the last decade, 
quantum analogs of these phases have been discovered in electron fluids.
Quantum-Hall systems and many strongly correlated electron
materials with very small structural anisotropy in the plane,
such as Sr$_{3} $Ru$_{2}$O$_{7}$, cuprates and pnictides, display very large anisotropies \cite{Kivelson:2003,Fradkin:2009,Yin:2010} 
in properties such as dc (Ref.~\onlinecite{Ando02}) and infrared conductivity,\cite{Lee02} 
spin fluctuations, \cite{Hinkov08} or Nernst signal.\cite{Daou10} 
Microscopic intraunit-cell nematicity in cuprates has also been observed.\cite{Kim:2010}

Much attention has been focused on whether nematic symmetry breaking in correlated electron materials, 
such as the cuprates, originates from the electron fluid or not. 
In fact, symmetry breaking in the electron fluid necessarily implies symmetry breaking in the structure and vice versa.\cite{Fradkin:2009} 
Recent studies show that a small band-structure anisotropy can lead to amplification of the anisotropy in low-energy excitations, 
even when spontaneous nematic symmetry breaking does not occur in the electron fluid. 
Slave-boson mean-field theory \cite{Yamase06} calculations, for example, 
suggest that the anisotropy in the magnetic excitation spectrum \cite{Hinkov08} can be enhanced by the
proximity to a correlation-induced Pomeranchuk instability. 
Another recent paper proposes an enhancement of anisotropy due to relaxation of frustration on the Kagome lattice 
which originates from the renormalization of the band structure near the correlation-induced metal-insulator transition, 
known as the Mott transition.\cite{Furukawa:2010}
Both these studies focus on the renormalized shape of the Fermi surface or on the electronic dispersion relation.

Using quantum cluster methods, here we find instead that a small orthorhombic distortion of hopping
in a model for cuprate superconductors leads to a large nematic anisotropy of the 
self-energy at small doping and large enough interaction strength.
The large anisotropy that we find also decreases when frequency is too large. 
For all these reasons the name dynamical nematicity is appropriate. 
The self-energy anisotropy dominates over the Fermi-surface deformation and appears in the pseudogap regime \cite{Ando02}
at low temperature as a finite-doping signature of the Mott transition.\cite{Sordi:2010} 
No additional spontaneously generated one-dimensional structures that could also break translational symmetry, such as stripes, are needed. 
This explains why the anisotropy that is easily observed in transport quantities is often hard to associate to static
stripes (as in YBa$_{2}$Cu$_{3}$O$_{y}$), even though such stripes are observed in some materials.\cite{Kivelson:2003} 
Our results are in qualitative agreement with angle-resolved photoemission spectroscopy (ARPES) 
(Ref.~\onlinecite{Fournier:2010}) and conductivity measurements \cite{Ando02} in highly underdoped YBa$_{2}$Cu$_{3}$O$_{y}$. 
There, large conductivity anisotropies are observed without major contributions from the CuO chains \cite{Ando02} 
and at temperatures above the appearance of quasielastic incommensurate peaks. \cite{Haug:2010}

We consider the two-dimensional Hubbard Model as the simplest one that exhibits the physics of electronic correlations 
\begin{equation}
H=-\sum_{ij\sigma }t_{ij}d_{i\sigma }^{\dag }d_{j\sigma} + U\sum_{i} d_{i
\uparrow}^\dag d_{i \uparrow} d_{i \downarrow}^\dag d_{i \downarrow}.
\label{eq:Hhub}
\end{equation}%
Here, $d_{i\sigma }$ is the annihilation operator for an electron with spin $\sigma $ at site $i$
and $U$ is the screened repulsive Coulomb interaction. 
The band structure part is described by $t_{ij}$ which includes the transfer integral for $i\neq j$ 
and the chemical potential $\mu $ for $i=j$. 
The next-nearest-neighbor transfer integral is $t^{\prime }$ and the band anisotropy is introduced 
via the nearest-neighbor transfer integral $t$ along the $x$ and $y$ directions as $t_{x,y}=t(1\pm \delta _{0}/2)$. 
We neglect small kinematic effects associated with anisotropy in lattice constants and take
orthorhombicity into account only dynamically through hopping.

We employ the cellular dynamical-mean-field theory (CDMFT) 
(Refs.~\onlinecite{Kotliar01} and \onlinecite{Kotliar06}) and the dynamic cluster approximation (DCA) 
(Refs.~\onlinecite{Hettler:dca} and \onlinecite{Maier05}) at zero temperature $T$; these methods are capable of
capturing the full dynamics (i.e., the frequency dependence of the spectral function) 
and the short-ranged spatial correlations beyond the single-site dynamical-mean-field theory. 
Both the CDMFT and DCA map the bulk lattice problem onto an effective Anderson model 
describing a cluster embedded in a bath of noninteracting electrons. 
The short-ranged dynamical correlations are treated exactly within the cluster. 
In DCA the embedded cluster is, in general, translationally invariant but not in CDMFT.

In this study, the $2\times 2$ plaquette coupled to bath orbitals is parameterized as
\begin{eqnarray}
&&\hspace{-1.5em}H_{clust}=-\sum_{ij=1}^{4}\sum_{\sigma
}t_{ij}^{c}d_{i\sigma }^{\dag }d_{j\sigma }+U\sum_{i=1}^{4}d_{i\uparrow
}^{\dag }d_{i\uparrow }d_{i\downarrow }^{\dag }d_{i\downarrow }  \notag \\
&&\hspace{-2.2em}+\sum_{\alpha =1}^{N_{\alpha }}\sum_{\mathbf{K}\sigma }%
\Biggl\{\varepsilon _{\mathbf{K}}^{\alpha }c_{\alpha \mathbf{K}\sigma
}^{\dag }c_{\alpha \mathbf{K}\sigma }+\sum_{i=1}^{4}\bigl(v_{\mathbf{K}%
i}^{\alpha }d_{i\sigma }^{\dag }c_{\alpha \mathbf{K}\sigma }+h.c.\bigr)%
\Biggr\}.  \label{eq:Hclust}
\end{eqnarray}
The first two terms describe the interacting sites on the cluster. 
For CDMFT, $t_{ij}^{c}=t_{ij}$ while for DCA, $t_{ij}^{c}=\frac{4}{\pi }t_{ij}$
for nearest-neighbor hoppings and $t_{ij}^{c}=\frac{16}{\pi ^{2}}t_{ij}$ for
second-neighbor hoppings.\cite{Maier05} 
In both cases, $t_{ii}^{c}=\mu $ and the interaction term is the same as in Eq.~(\ref{eq:Hhub}). 
The third and the fourth terms represent bath-orbital levels and cluster-bath hybridizations, respectively. 
Bath orbitals ($\alpha $) are classified by the cluster momenta $\mathbf{K}=(0,0),(\pi ,0),(0,\pi ),(\pi ,\pi )$ 
(Refs.~\onlinecite{Haule07} and \onlinecite{Liebsch08}) and 
the impurity-bath hybridization reflects the symmetry as 
$v_{\mathbf{K}i}^{\alpha }=v_{\mathbf{K}}^{\alpha }e^{i\mathbf{K}\cdot \mathbf{r}_{i}}$. 
The Anderson model, Eq.~(\ref{eq:Hclust}), is solved using the Lanczos exact diagonalization technique.\cite{Caffarel94,Capone:2004,Kancharla08} 
This technique limits the number of bath orbitals ($N_{\alpha }=2$) but allows us to access the dynamical 
quantities directly on the Matsubara, $i\omega _{n}$, or real-frequency axis $\omega +i\eta $.

The Anderson model is subjected to a self-consistency condition, relating its one-particle cluster Green's function 
$G_{ij}^{imp}(i\omega_{n})=\,-\int d\tau e^{i\omega _{n}\tau }\,\langle T_{\tau }d_{i}(\tau) d_{j}^{\dag }(0)\rangle $ 
to the lattice Green's function of the original model as
\begin{eqnarray}
\hat{G}^{imp}(i\omega _{n}) &=& N_c \int \biggl(\frac{d\tilde{k}}{2\pi } %
\biggr)^2 \hat{G}(\widetilde{\mathbf{k}}, i \omega _{n}),  \label{selfcon} \\
\hat{G}(\widetilde{\mathbf{k}}, i \omega _{n}) &=&\left[ \left( i\omega
_{n}+\mu \right) \hat{I}-\hat{t} (\widetilde{\mathbf{k}})- \hat{\Sigma}%
(i\omega _{n})\right] ^{-1},
\end{eqnarray}
where $\hat{\Sigma}(i\omega _{n})=\hat{G}_{U=0}^{imp}(i\omega _{n})^{-1}-\hat{G}^{imp}(i\omega _{n})^{-1}$. 
Here, ${t}_{ij}(\widetilde{\mathbf{k}})=N_{c}^{-1}\sum_{\mathbf{K}}
e^{i\left( \mathbf{K}+\vartheta \widetilde{%
\mathbf{k}}\right) \cdot \left( \mathbf{r}_{i}-\mathbf{r}_{j}\right)
}\varepsilon _{\mathbf{K+}\widetilde{\mathbf{k}}}$ 
describes the hopping between the clusters covering the original lattice with $\vartheta =1$ for
CDMFT and $\vartheta =0$ for DCA,\cite{Maier05} $\widetilde{\mathbf{k}}$ are
wave vectors in the reduced Brillouin zone, $\varepsilon _{\mathbf{k}%
}=-2(t_{x}\cos k_{x}+t_{y}\cos k_{y}+2t^{\prime }\cos k_{x}\cos k_{y})$ 
the non-interacting dispersion relation of the lattice, and $N_{c}=4$ for our $2\times 2$ plaquette. 
The self-consistency Eq.~(\ref{selfcon}) is solved by iteration, recomputing the bath Green's function $\hat{G}_{U=0}^{imp}$, 
and thus new bath parameters, at each iteration by minimizing with a conjugate gradient algorithm a distance function that includes
frequency dependence on the discrete Matsubara frequency $\omega_{n}=(2n-1)\pi /\beta $, 
with the small energy cut-off $\beta t=50$, a frequency weighting $\left\vert 1/\omega _{n}\right\vert $, 
and a large frequency cut-off of $10t$.\cite{Capone:2004,Senechal:2010}

The cluster self-energy $\hat \Sigma \left(i \omega_n \right)$ enters a CDMFT lattice Green's function 
$\hat{G} (\widetilde{\mathbf{k}}, i \omega_n)$ 
with larger unit cell. From the self-energy functional point of view\cite{Potthoff:2003} 
it is natural to focus on diagonal quantities. 
This corresponds to Green's function periodization 
$G(\mathbf{k}, i \omega_{n})=\,N_{c}^{-1}\,\sum_{ij}\,e^{-i\mathbf{k}\cdot (\mathbf{r}_{i}-\mathbf{r}%
_{j})} \,G_{ij}(\mathbf{k}, i \omega _{n})$.\cite{Senechal00}
Other periodization schemes have been proposed such as cluster cumulant \cite{Stanescu06} and self-energy \cite{Biroli02} periodizations. 
In DCA, the conductivity can be computed including vertex corrections and 
without restoring the translational lattice invariance of the self-energy.\cite{Lin:2010} 
We will show that the results we find with CDMFT and Green's function periodization are qualitatively consistent with DCA. 
This validates our approach. 
In CDMFT, a general procedure for vertex corrections has not been developed yet.

\begin{figure}[tbp]
\includegraphics[width=0.9\columnwidth,clip]{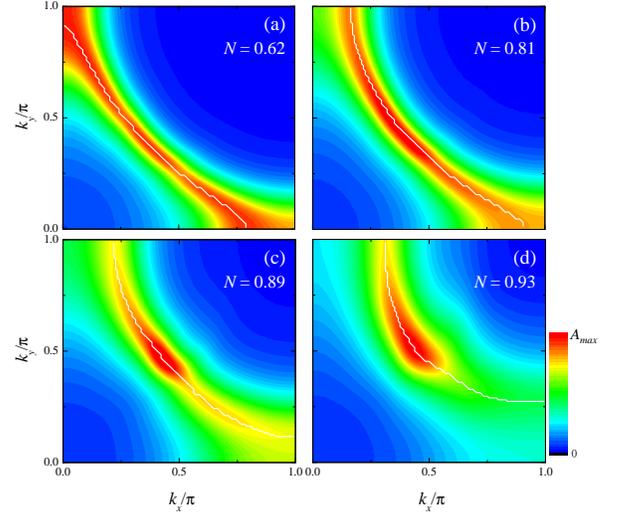}
\caption{(Color online) Spectral function at the Fermi level $A(\mathbf{k},%
\protect\omega =0)$ in the first quadrant of the Brillouin zone obtained
with CDMFT, $U=10$, $t^{\prime }=-0.3t$. The self-energy is computed
directly in real frequency with a small imaginary part $i\protect\eta $ with
$\protect\eta =0.1t$. The carrier density is (a) $N=0.62$, (b) $0.81$, (c) $%
0.89$ and (d) $0.93$. The maximum of the spectral intensity $A_{max}$ is
given by (a) 0.9, (b) 0.6, (c) 0.4, and (d) 0.3 in units of $1/t$. The thin
white lines indicate the location of the maxima of $A(\mathbf{k},\protect%
\omega =0)$ that define the Fermi surface. The band anisotropy is $\protect%
\delta _{0}=0.04$. }
\label{MDC}
\end{figure}

We use $t^{\prime }=-0.3t$ and $U=10t$, as parameters that are appropriate for cuprates. 
The $U$ dependence of the results for the conductivity anisotropy will be given to illustrate the physics. 
We take as intrinsic orthorhombic anisotropy in the nearest-neighbor hopping $\delta_{0}=0.04$ 
which is small, like in most cuprates.\cite{Ando02,Yamase06} 
For $\delta _{0}=0$, in agreement with previous quantum cluster calculations,\cite{Gull09} 
we do not find spontaneous breaking from $C_{4}$ to $C_{2}$ symmetry. 
In other words, we do not find a nematic instability. 
This kind of spontaneous symmetry breaking is also often referred to as the Pomeranchuk instability or effect. 
It is possible that in larger clusters, spontaneous symmetry breaking occurs at the van Hove doping, 
as found at weak coupling.\cite{Halboth00} 
Indeed, near this doping, ($N=0.727$ at $\delta _{0}=0)$ we had difficulty in obtaining well-converged solutions 
when $\delta _{0}\neq 0$.

Figure~\ref{MDC} illustrates the strong anisotropy in $A(\mathbf{k},\omega=0)=-\frac{1}{\pi }\mathrm{Im}G (\mathbf{k},\omega =0)$ 
that emerges as one approaches half filling. 
The results are displayed in the first quadrant of the Brillouin zone. 
The Fermi surface, defined as the location of the maxima of $A(\mathbf{k},\omega =0)$, is shown as a thin white line in Fig. 1. 
It is not strongly anisotropic, in other words the real part of the self-energy is not strongly affected. 
On the other hand, if we define overdoping as $N < 0.85$ and underdoping as $N>0.85$, \cite{Kancharla08} 
then in the latter regime the spectral \textit{intensity} shows strong directional anisotropy;
at $N=0.93$, for example, the spectral weight near $(\pi ,0)$ is suppressed by as much as $54\%$ compared with that near $(0,\pi)$. 
In other words, the pseudogap that normally appears symmetrically near $\left( \pi ,0\right)$ and $\left( 0,\pi \right) $ 
is now much more strongly anisotropic than the $4\%$ band anisotropy. 
This reflects a strong anisotropy in the self-energy. 
Recent ARPES experiments on YBa$_2$Cu$_3$O$_y$ have shown similar anisotropy in the lightly doped regime.\cite{Fournier:2010}

\begin{figure}[tbp]
\includegraphics[width=0.8\columnwidth]{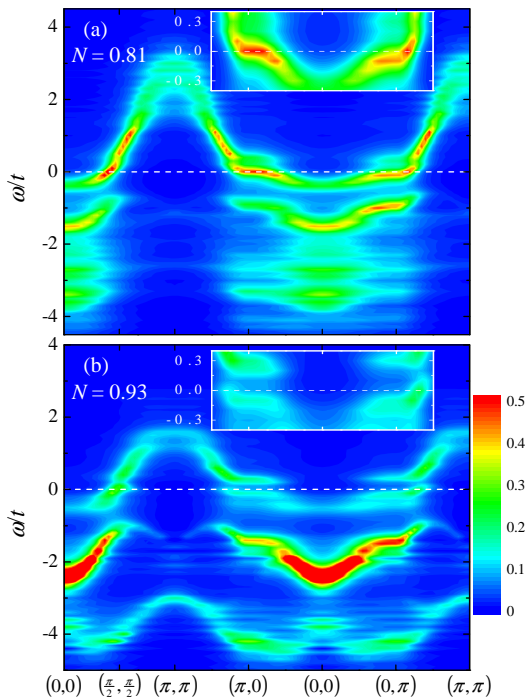}
\caption{(Color online) Angle-resolved spectral function $A(\mathbf{k},%
\protect\omega )$ near the Fermi level obtained from CDMFT with the same
parameters as Fig.1. (a) $N=0.81$ and (b) $N=0.93$. Spectral intensity is
given in units of $1/t$. The Fermi level $\protect\omega =0$ is indicated by
a broken line. In the insets, the vertical axis near the Fermi level is
blown up.}
\label{arpes}
\end{figure}

Let us now turn to the finite frequency regime. 
In Fig.~\ref{arpes}, we present the angle-resolved spectral function $A(\mathbf{k},\omega )$ 
for the overdoped (a) and the underdoped (b) regimes. 
The blow up of the region near the Fermi level in the insets shows that the anisotropy is most pronounced
in the underdoped regime, $N=0.93$. 
The anisotropy also extends to progressively higher frequencies as the system is underdoped. 
The strong anisotropy essentially disappears at high frequency, reminiscent of recent observations on the spin-fluctuation spectrum.\cite{Hinkov08} 
This indicates that the nematicity here is a dynamical phenomenon associated with the electronic response near the Mott transition as $N\rightarrow 1$.

\begin{figure}[tbp]
\includegraphics[width=0.8\columnwidth,clip]{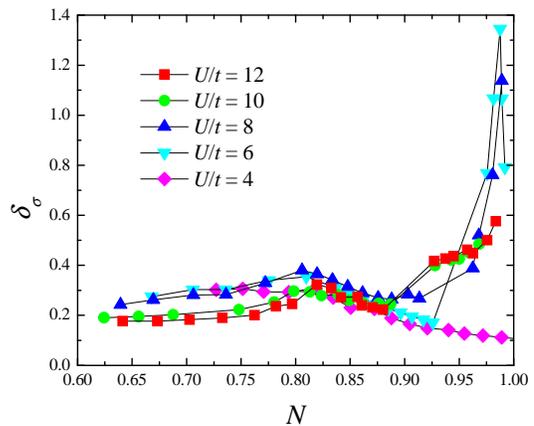}
\caption{(Color online) Anisotropy in the CDMFT conductivity $\protect\delta %
_{\protect\sigma }=2\left[ \protect\sigma _{x}(0)-\protect\sigma _{y}(0)%
\right] /\left[ \protect\sigma _{x}(0)+\protect\sigma _{y}(0)\right]$ as a
function of filling $N$ for various values of $U$ and $\protect\eta =0.1 t$, 
$\protect\delta _{0}=0.04$.}
\label{anisotropy_CDMFT}
\end{figure}

The link with Mott physics is seen most clearly in the anisotropy of the dc conductivity $\sigma _{x(y)}$ that we consider now. 
For the CDMFT results, we use the Kubo formula with the periodized Green's function to compute
\begin{equation}
\sigma _{x(y)}(\omega =0)=\frac{2e^{2}}{\pi \hbar }\int \biggl(\frac{dk}{%
2\pi }\biggr)^{2}\biggl(\frac{\partial \varepsilon _{\mathbf{k}}}{\partial
k_{x(y)}}\biggr)^{2}\bigl[\mathrm{Im}G(\mathbf{k},0)\bigr]^{2}.
\end{equation}
This formula neglects vertex corrections. 
The conductivities $\sigma _{x(y)}$ extrapolate to infinity like $1/\eta $ as the small imaginary part $\eta$
tends to zero, as expected in a pure metal (Drude peak) at $T=0$. 
We have checked however that $\sigma _{y}/\sigma _{x}$ reaches a limit. 
Nevertheless, we use $\eta =0.1 t$ in Fig.~\ref{anisotropy_CDMFT} to avoid uncertainties in extrapolating the conductivity anisotropy 
$\delta _{\sigma}=2\left[ \sigma _{x}(0)-\sigma _{y}(0)\right] /\left[ \sigma _{x}(0)+\sigma_{y}(0)\right]$. 
With smaller $\eta$ the anisotropy is generally larger.

As a function of filling $N$ and for all $U$, there is a peak around $N\sim0.8$, 
not far from the non-interacting $\delta _{0}=0$ van Hove singularity at $N=0.727$. 
This is a band-structure effect since the Fermi surface opens up as if the system had a tendency to be quasi-one-dimensional 
[see Fig.~\ref{MDC} (b)]. 
We found little $\eta $ dependence of $\delta _{\sigma }$ near $N=0.8$ 
because of the large imaginary part of the cluster self-energy for that filling.\cite{Haule07}

The most interesting results occur close to half filling. 
For values of $U$ below the critical $U_{c1}$ for the Mott transition,\cite{phk} 
the anisotropy takes small values consistent with the small orthorhombic distortion. 
At $U=6 t$, the anisotropy in conductivity is largest and not monotonic close to $N=1$. 
This $U$ is just slightly above the critical $U_{c1}\sim 5.25 t$ for the Mott transition obtained for CDMFT 
with exact diagonalization and $t^{\prime }=0$.\cite{balzer} 
The abrupt and large increase of the anisotropy at a finite doping close to half filling 
for all larger values of $U$ is similar to the experiment.\cite{Ando02} 
The CDMFT results are consistent with the existence of a first-order transition
between two kinds of metal found recently for $U$ larger than $U_{c2}$ at finite doping.\cite{Sordi:2010} 
The metal closest to half filling displays a pseudogap and we have shown in Figs. \ref{MDC} and \ref{arpes} that in this
metallic phase it is extremely sensitive to small orthorhombic distortions.
The anisotropy in occupancy of $\mathbf{K}$ orbitals, at most 12\%, does not suffice to explain the anisotropy in conductivity. 
The imaginary part of the self-energy plays a prominent role since replacing it
by a constant in the optical conductivity formula removes the large anisotropy.
The fact that the enhanced anisotropy occurs far from the van Hove singularity
suggests that it is not simply a weak-coupling effect like the Pomeranchuk instability.\cite{Halboth00}

\begin{figure}[tbp]
\includegraphics[width=0.8\columnwidth,clip]{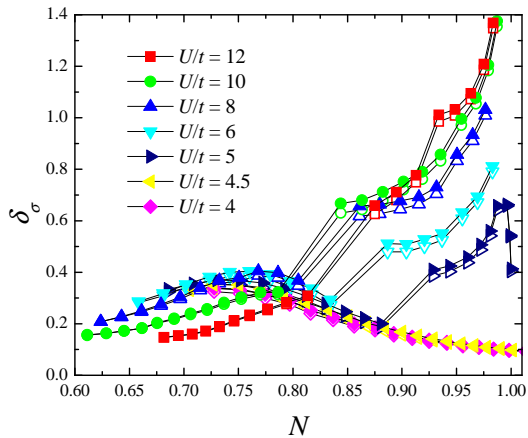}
\caption{(Color online) Anisotropy in the DCA conductivity for $\protect%
\delta _{0}=0.04$ as a function of filling $N$ for various values of $U$ and
$\protect\eta =0.1 t$. Open symbols do not include vertex corrections, filled
symbols do. }
\label{anisotropy_DCA}
\end{figure}

The above results are confirmed by Fig.~\ref{anisotropy_DCA}, obtained from DCA calculations 
that include vertex corrections.\cite{Lin:2010,vertex,DCA_4*} 
The value of $U_{c}$ is about $4.5 t$ for DCA at $t^{\prime}=0$.\cite{Gull:2008} 
It is above this value that the large anisotropies are present. 
The critical filling where the anisotropy increases does not depend strongly on $\delta _{0}$. 
Anisotropy scales like $\delta _{0}$ at small $\delta _{0}$ and is linked with the pseudogap 
so it should appear at comparable $T$.\cite{Macridin:2005} 
Other interesting physics has been found in DCA at $U>U_{c}$,\cite{Khatami:2010,Gull:2010} 
including a critical filling similar to ours at $U=8 t$
beyond which a local bond-order susceptibility diverges at $T\rightarrow 0$.\cite{Macridin:2008}

We have shown that, in the presence of interactions larger than the critical $U$ for the Mott transition, 
small orthorhombicity can lead to very large anisotropies in the pseudogap and conductivity in the underdoped regime
close to half filling without further symmetry breaking. 
Transport anisotropy should be much more sensitive to uniaxial pressure in the underdoped than in the overdoped regime. 
Mott-induced dynamical nematicity should influence all other transport properties. 
The competition between this effect and other symmetry-broken phases would be worth investigating. 
Preliminary study reveals that, in the presence of small band anisotropy, the superconducting order parameter becomes highly
anisotropic, $\Delta _{x}\neq -\Delta _{y}$,\cite{Civelli} reminiscent of experiment.\cite{Lu:2010}

We thank G. Sordi, J. Chang, and L. Taillefer for discussions. 
The work of S.O. was supported by the Materials Sciences and Engineering Division, 
Office of Basic Energy Sciences, U.S. Department of Energy. 
This work was partially supported by NSERC (Canada) and by the Tier I Canada Research Chair Program (A.-M.S.T.). 
Some of the computational resources were provided by RQCHP and Compute Canada.

\end{document}